\documentclass[letterpaper, 10 pt, conference]{ieeeconf}  % Comment this line out if you need a4paper
\usepackage{amsmath,amssymb,amsfonts}
\usepackage{mathtools}
\usepackage{algorithm}
\usepackage{algpseudocode}
\usepackage{array}
\usepackage{textcomp}
\usepackage{multirow}
\usepackage{stfloats}
\usepackage{xcolor}
\usepackage{url}
\usepackage{verbatim}
\usepackage{nameref}
\usepackage{graphicx}
\usepackage{ulem}

\usepackage{dsfont}
\usepackage{cite}
\usepackage[colorlinks=true, linkcolor=blue, citecolor=blue, urlcolor=blue]{hyperref}
\usepackage{wrapfig} 
\usepackage{listings}

\usepackage{tabularx}
\usepackage{booktabs}
\usepackage[flushleft]{threeparttable}
\usepackage{makecell}
\usepackage{multirow}
\usepackage{graphicx}
\usepackage{url}

\newcommand{\gaby}[1]{}

\IEEEoverridecommandlockouts                              % This command is only needed if 
\title{\LARGE \bf
Persistent LLM Orchestration of Heterogeneous Robot Teams via Partial Delegation to Multi-Agent Protocols}

\author{Maria G. Mendoza$^{1,*}$, Victoria M. Tuck$^{2,*}$, Zachary Ravichandran$^{2}$, Shankar Sastry$^{1}$, and George J. Pappas$^{2}$%
\thanks{$^{*}$Equal contribution and corresponding authors: {\tt\small maria\_mendoza@berkeley.edu}, {\tt\small vtuck@seas.upenn.edu}.}%
\thanks{$^{1}$Department of Mechanical Engineering (M.G.M.) and Department of Electrical Engineering and Computer Sciences (S.S.), University of California, Berkeley, CA, USA.}%
\thanks{$^{2}$Department of Electrical and Systems Engineering, University of Pennsylvania, Philadelphia, PA, USA.}%
}

\begin{document}

\maketitle
\thispagestyle{empty}
\pagestyle{empty}

%%%%%%%%%%%%%%%%%%%%%%%%%%%%%%%%%%%%%%%%%%%%%%%%%%%%%%%%%%%%%%%%%%%%%%%%%%%%%%%%
\begin{abstract}
Natural language specifications let non-expert users task robot teams, and large language models (LLMs) have recently proven effective at grounding such semantic specifications into plans for heterogeneous teams. Existing approaches assume missions that terminate and environments that, while unknown, do not change, but both assumptions fail in many emerging applications. For example, a disaster-response mission does not terminate, and its environment changes continuously as hazards spread and evacuees move. We identify three ways an LLM orchestrator breaks in this setting: its prompt grows without bound, its reasoning degrades as agents and targets accumulate, and it stops actively monitoring after covering the search space once. We build on the successful framework of grounded generative intelligence, and address each issue with a classical idea. A Markovian state summary, rebuilt from the semantic map at every invocation, replaces the growing interaction history. Partial delegation lets the orchestrator assign a subset of robots to an autonomous multi-agent protocol and reserve its reasoning for generating subtasks that require detailed parameters and careful ordering. Lastly, observation staleness, surfaced explicitly in the prompt, together with task- and event-triggered re-prompting, keeps the team vigilant. In a simulated urban search-and-rescue mission with spreading fire and moving evacuees, our framework rescues 98\% of civilians versus 89\% for the strongest baseline while using six times fewer tokens. Additional ablations isolate the contribution of each mechanism.

\end{abstract}

%%%%%%%%%%%%%%%%%%%%%%%%%%%%%%%%%%%%%%%%%%%%%%%%%%%%%%%%%%%%%%%%%%%%%%%%%%%%%%%%
\gaby{Comments from PY: "I feel that the message of the first paragraph is not very clear. The beginning seems to suggest that multi-robot approaches are good. Then it jumps to an example and then describes how the LLM can help. Perhaps splitting into two paragraphs would help. For the related work section, I would suggest adding one sentence after each paragraph to describe how your work is different from previous work"}
\section{INTRODUCTION}
Multi-robot approaches offer several advantages over single-robot systems: they scale to larger environments, provide redundancy against individual failures, and let simpler robots combine for complex tasks. These properties are useful across a range of application domains including infrastructure maintenance, healthcare robotics, and  disaster response, where specifically the danger of the scenario makes offloading tasks to robots especially valuable. Consider evacuation during a wildfire: rapid response is required, resources are limited, the environmental state changes continuously, humans may act irrationally, and, critically, the mission has no terminal ``done'' signal---it is continuous. The first responders directing the team are not roboticists, and their needed tasks cannot be enumerated in advance, For example, "check whether anyone is still on Main Street" is not a coverage objective or a fixed task list. 
% Natural language is the interface such users already have, and LLMs have recently made it a practical one for tasking heterogeneous teams. Additionally, the first responders may not be robotics experts and will therefore need to interact with the system using a more familiar medium.
Robotic tasks must therefore be not just allocated but also \textit{planned}.
Due to the recent success in generative intelligence such as large language models (LLMs) and vision-language models (VLMs), natural language has gained interest as a medium for task planning and allocation for these \textit{semantic} mission specifications. 

However, prior work in LLM-based multi-robot allocation makes two assumptions that do not hold for such continuous missions in time-varying settings: (1) tasks can be cleanly labeled as finished or unfinished, and (2) the environment, while possibly unknown, is exogenously static. Additionally, for continuous missions, scalability becomes a more relevant concern due to increasing time on top of many agents, but LLMs have already been shown to struggle with reasoning as the number of agents \cite{chen2024scalable} and context size \cite{generative-agents} grow. In this work, we aim to relax both assumptions and improve the scalability of LLM-based orchestration of multi-robot systems in order to effectively handle continuous missions in evolving environments. These considerations lead to our motivating question:

\vspace{2mm}
\textit{For continuous, semantic missions in unknown, time-varying, and time-critical environments, how do we efficiently and persistently coordinate a heterogeneous team of robots?}
\vspace{-2mm}

Our answer is to let the LLM orchestrate both robots and algorithms. We build on prior work in grounded LLM-based task planning and allocation for heterogeneous teams \cite{ravichandran2025heterogeneous}, and introduce three ideas aimed at efficiency and persistence. First, 
% standard re-prompting strategies pass either the entire interaction history or a fixed window. Due to large context windows, passing the full history is typically not an issue, but a large context is expensive under constrained resources and grows without bound as the mission continues. Instead, we show that 
the orchestrator is provided a Markovian summary of the state of the system and its tasks, updated from the map at every iteration, instead of the typical full history so that prompt size does not grow unbounded with time.
% maintains the success rate while significantly reducing token cost.
Second, through \textit{partial delegation} the orchestrator assigns a subset of robots to an autonomous multi-agent protocol such as ergodic coverage [6], which runs continuously without further prompting, allowing the LLM to reserve reasoning for the semantically challenging tasks of subtask generation, parameter choice, and determining task precedence. 
% that remainWe show that our system scales better with the number of agents when delegating a portion of the agents to these already existing multi-agent protocols in order to reduce the number of tasks necessary for the LLM to plan.
Third, to invoke vigilance in the time-varying environment, the Markovian state exposes observation staleness explicitly, and the orchestrator is re-invoked on task completion, task failure, and relevant map changes.
% in a time-varying setting, it is not enough to trigger on subtask completions, which are triggered too infrequently or on observation changes, which can cause stalls if the observation is not continuously tasked. Instead, we leverage a mixed approach, which we compare to an MPC-style trigger at a fixed frequency and baseline event-based or task-based triggers that fire on observation changes and task completions, respectively. We show that our approach balances efficiency of task assignment while maintaining continuous operation.

\begin{figure}
    \vspace{0.2cm}
    \centering
    \includegraphics[width=1.0\linewidth]{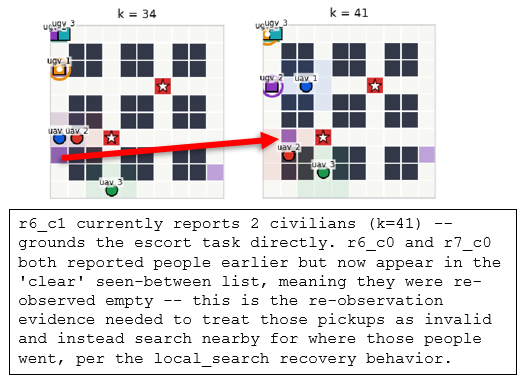}
\caption{The orchestrator reasoning about moving humans under partial
information. Left: at $k=34$ people are reported in \texttt{r6\_c0} and \texttt{r7\_c0}, with an escort under way to \texttt{r7\_c0}. Right: by $k=41$ both regions have been re-observed clear and two civilians are reported in \texttt{r6\_c1}. The orchestrator treats the re-observation as evidence the pickups are stale, cancels the escort, issues a \texttt{local\_search} around each vacated region to recover the missing group, and escorts the live report at \texttt{r6\_c1}.}
    \label{fig:operation}
\end{figure}

\textbf{\textit{Our contributions are}}: (1) an LLM orchestration framework for continuous multi-robot missions in unknown, time-varying environments, built on three mechanisms: a Markovian state summary, partial delegation to multi-agent protocols, and staleness-aware task- and event-triggered re-prompting (Table \ref{tab:complexities}). (2) An empirical study in a simulated urban search-and-rescue mission with spreading fire and moving evacuees showing 98\% rescue at one-sixth the token cost of the strongest baseline (Sec. \ref{sec:results}). (3) Case studies and ablations showing why each mechanism is needed: a Markovian representation of history reduces token usage and improves percent success (Table \ref{tab:results}), delegation frees the orchestrator to reason more successfully and efficiently about ambiguous, moving targets (Fig.~\ref{fig:operation}), and exposed staleness and varied triggering are what make the team persistently revisit regions it has already covered (Fig. \ref{fig:data_age}).

\begin{figure*}[t]
    \centering
    \includegraphics[width=\linewidth]{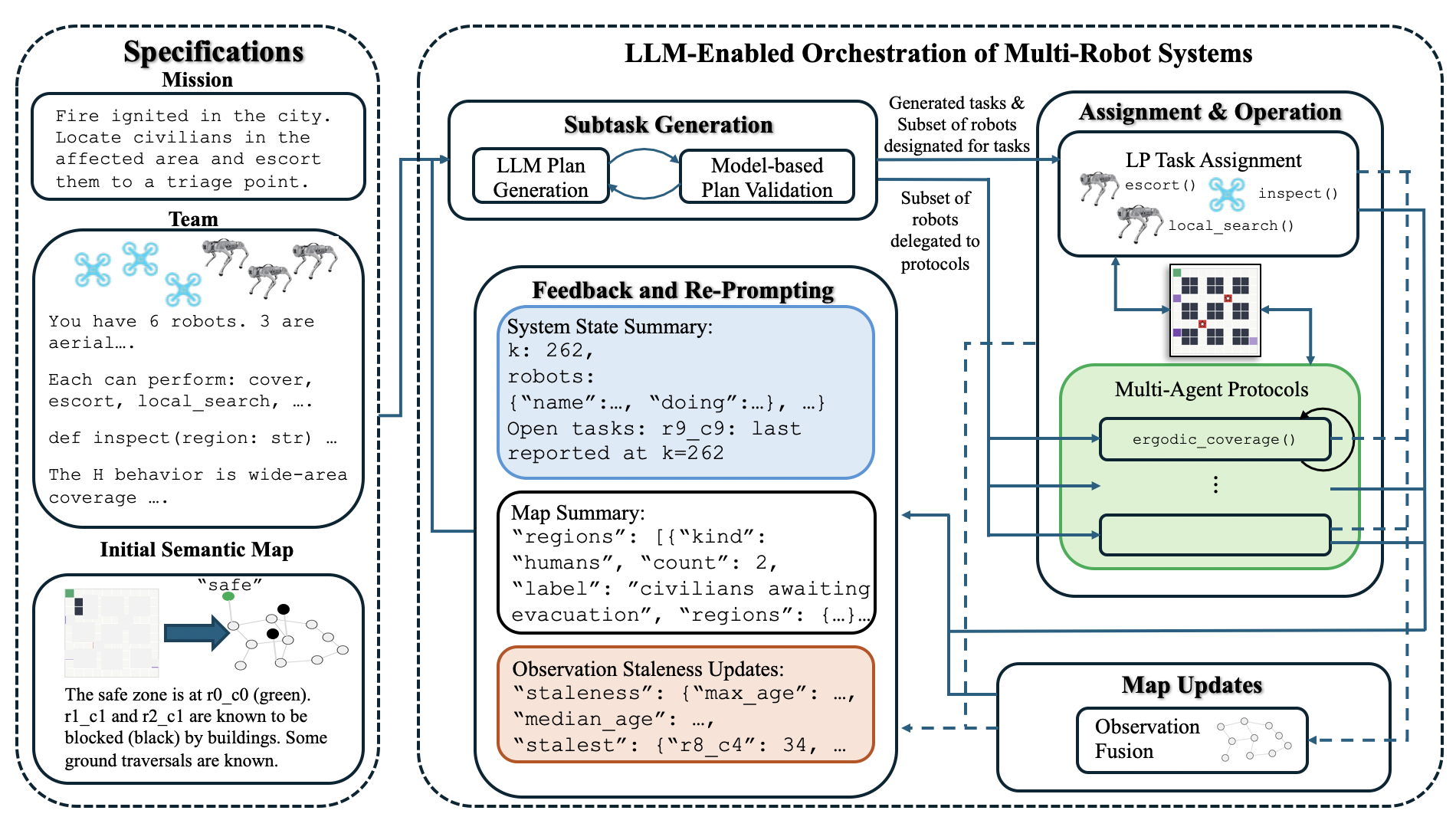}
    \caption{System architecture. The mission statement, team specification, including task primitives and multi-agent protocols, and a partial semantic map are provided as inputs. The LLM is prompted with these inputs and the prompt information of Sec. \ref{subsec:dispatcher} to delegate a subset of the robots to the multi-agent protocols and generate subtasks for the rest according to the system needs. This plan is validated, and the LLM is provided feedback if there are issues. The actual task assignment is done symbolically with a linear program (LP) assignment. Multi-agent Protocols (green) are shown to be continuously operating with the self-loop or similar to tasks in that they have an endpoint. Asymmetric observations from all operations are constantly fused into the map. The feedback mechanism is triggered on task completion (solid lines) or on relevant map update events (dashed lines). Feedback on the map and observation age (orange) is compiled with a state summary of the history (blue) and provided in the new prompt to the LLM.}
    \label{fig:architecture}
\end{figure*}
\section{RELATED WORKS}

\noindent  \textbf{Multi-robot coordination in time-varying environments}
Effectively coordinating multi-robot teams in dynamic environments is an outstanding challenge. While there is extensive work on dynamic multi-robot task allocation \cite{CHAKRAA2023104492, mendoza2026dynamic, NFM_2024}, the combined problem of dynamic task arrival and time-varying environments is limited. Along with the dynamic nature of the environment, other factors may be considered such as uncertainty, temporal constraints in the mission specification, or payload constraints~\cite{choudhury2021dynamic, ghassemi2022multirobot, Liu2026LearningAO, chen2019distributed}.

\noindent  \textbf{LLMs for Multi-Robot Coordination.} Recent work uses LLMs to translate natural-language mission specifications into multi-robot coordination. 
Some approaches use LLMs to frame a reinforcement learning problem that solves some mission specification~\cite{sebastian2026prompting}.
Other approaches directly use the LLM as a high-level planner by providing the model with a set of high-level robot behaviors (e.g., for navigation, manipulation), and then tasking it with providing a plan for each robot~\cite{chen2024scalable, mandi2023roco}.
Such approaches may operate in a closed-loop manner, where plans are updated as robots gather feedback~\cite{kannan2024smart}.
The LLM-as-coordinator approach has also been adopted for heterogeneous robot teams, which require the LLM to reason about robot-specific capabilities~\cite{liu2025coherent, tfrairground}.
One challenge of using LLMs in-the-loop for planning is their tendency to hallucinate. 
Several works address this challenge with architectures that combine LLMs with non-learned components, such as optimization-based task allocation or conformal prediction for uncertainty quantification~\cite{wang2025probabilistically}.
These approaches ground LLM outputs in verifiable representations, which all require plans to be vetted before execution~\cite{ravichandran2025heterogeneous, jiaxing2026tool, zhu2025dexterllmdynamicexplainablecoordination, strader2025languagegroundedhierarchicalplanningexecution}.
Moreover, the existing literature suggests that centralized planning approaches exhibit better scaling properties in complex environments, as compared to decentralized approaches~\cite{ravichandran2025heterogeneous, liu2025coherent, kannan2024smart, strader2025languagegroundedhierarchicalplanningexecution}.

\noindent \textbf{Closed-loop planning with LLMs} Effectively incorporating feedback is crucial when planning in dynamic and time-varying environments~\cite{bhat2024grounding}. 
Event-driven planning mechanisms, wherein robots provide feedback from onboard perception on task execution to a central LLM planner, have been proposed in \cite{borate2026comuros}.
Semantic graph updates have also been shown to be an effective source of feedback for planning in partially known environments~\cite{spine-Ravichandran-2025}.
The use of structured representations, such as behavior trees, make for effective sources of feedback and validation~\cite{huang2025compositional}

\section{PROBLEM DESCRIPTION}
\label{sec:problem}

We focus on the problem of multi-robot task planning and allocation for a continuous mission in a time-varying environment. An LLM planner has access to M robots with possibly heterogeneous types and a semantic description of the mission. Each robot type has a set of capabilities, and some or all capabilities may overlap between different robot types. The environment may have regions that are restricted to certain robot types, such as a dense forest that cannot be traversed by a ground robot but can be viewed by an aerial vehicle. We assume the LLM knows the bounds of the environment but little about the structure or contents, time-varying or static. The LLM planner is provided with textual descriptions for a set of task primitives that can be assigned to agents as well as for a set of multi-robot protocols to which it can delegate a portion, all, or none of the robots. Each of the task primitives and protocols requires specific robot capabilities. For feasibility, we assume each of the regions can be viewed by at least one robot with sensing capabilities.

\section{APPROACH} \label{sec:approach}

We summarize our complete approach in Fig. \ref{fig:architecture}. Sec.~\ref{sec:mechanisms} details the changes required for this continuous, time-varying setting; Sec.~\ref{sec:map} describes the semantic map; Sec.~\ref{sec:primitives_protocols} presents task primitives and multi-agent protocols that ground generative intelligence to reduce errors; Sec.~\ref{subsec:dispatcher} covers the prompt structure and dispatch process; Sec.~\ref{sec:assignment} gives the optimization-based assignment; Sec.~\ref{sec:triggers} presents the re-prompting triggers; and Sec.~\ref{sec:feedback} details the feedback structures for persistence and reduced token usage.

\subsection{Mechanisms for Continuous Missions in Time-Varying Environments} \label{sec:mechanisms}
For semantic missions, LLMs are capable planners that produce reasonable subtask decompositions from a description of the environment and team \cite{ravichandran2025heterogeneous}. However, current methods struggle with continuous missions in time-varying environments. Table~\ref{tab:complexities} summarizes the mechanisms that we introduce to extend a naive re-prompting loop of previous approaches to efficient and persistent operation for continuous and time-critical missions in time-varying environments. In particular, for time-varying settings, the orchestrator must balance responding to changes (Time-varying environment) while actively monitoring for relevant changes (Mission persistence). Additionally, asymmetric sensing and noise between robots make data association ambiguous. For example, a far-away drone sees a human that a ground vehicle then goes to investigate; if it sees a human in a pink shirt in a slightly different spot, is that the same human?
Challenges with continuous missions arise from the accumulation of history (Scale, Memory) and the need for persistence of operation (Mission persistence). A long history in the prompt increases token usage, which is exacerbated by agent and map size, and can unnecessarily degrade the quality of the plan; persistence in discovery is a necessary trait in a non-terminating mission.
% the team's own actions alter it as well, \victoria{We don't cover this though we could if we want to include it in discussion or conclusion}

\begin{table}[t]
\vspace{0.2cm}
\caption{Failure modes and the proposed mechanisms}
\label{tab:complexities}
\centering
\small
\renewcommand{\arraystretch}{1.25}
\begin{tabularx}{\columnwidth}{@{}XX@{}}
\toprule
\textbf{Failure mode} & \textbf{Mechanism} \\
\midrule
Time-varying environment \newline
\textit{Exogenous, endogenous, and epistemic changes}
& Invocation triggered by changes to objects of interest (Sec. \ref{sec:triggers}) \\
\addlinespace
Memory  \newline
\textit{Accumulated history grows and describes outdated states, distracting the orchestrator}
& Markovian input, rebuilt from the current map at every invocation (Sec. \ref{subsec:dispatcher})\\
\addlinespace
Dynamic objects  \newline 
\textit{Asymmetric observation aliases distinct objects, making data association ambiguous}
& Pending tasks and failures trigger mechanism (Sec. \ref{sec:triggers})\\
\addlinespace
Scale \newline 
\textit{Planning degrades as team, targets, and map size grow}
& Delegation to a multi-robot protocol (Sec. \ref{sec:primitives_protocols}); assignment by linear
program (Sec. \ref{sec:assignment})\\
\addlinespace
Mission persistence \newline
\textit{The mission has no completion signal, and the orchestrator has no reason to revisit regions} & Observation age summarized in the state: mean age and the regions
observed least recently (Sec. \ref{sec:feedback})\\
\bottomrule
\end{tabularx}
\end{table}
\subsection{Shared Semantic Map} \label{sec:map}
We model the world via a topological graph-based \textit{semantic map} where nodes are regions or objects and edges are connections between regions denoting traversability or a connection between a region and an object denoting the existence of the object within that region. Each region is tagged with the time it was last observed. Each traversal edge includes a set of robot capabilities that allow traversal. The planner and agents have access to a partial initial semantic map called so because the objects and regions can carry semantic information such as "the tree is in the path". During operation, this semantic map may not always be accurate due to the addition of noisy observations.

\subsection{Task Primitives and Multi-Robot Protocols} \label{sec:primitives_protocols}

Like previous work in grounded generative intelligence, we provide our planner with a suite of \textit{task primitives} represented as a header with name, required inputs, if needed, and a description. An example primitive is shown in Fig. \ref{fig:task_primitive}. We also provide a suite of \textit{multi-agent protocols}. At each iteration, the planner decides only the number of agents to operate according to one of these protocols, which it can vary from iteration to iteration. An example protocol is ergodic coverage as shown in Fig. \ref{fig:ma_protocol}. These protocols can run autonomously and continuously even when the planner is re-prompted or chooses to adjust the number of robots running this protocol. This delegation decreases the specific tasks and task parameters the planner must reason about, leveraging the large body of work on multi-agent systems to improve scalability and reasoning.

\begin{figure}[t]
\begin{lstlisting}
def go_to(region: str) -> bool:
    """Navigate to `region'. This MUST be an existing region id.

    Positioning only -- it reports nothing. Use 
    `inspect' if you want to know what is there.  Reach for this when the value is being somewhere."""
\end{lstlisting}
\caption{An example task primitive for the \lstinline[language=Python]|go_to()| region task.}
\label{fig:task_primitive}
\end{figure}

\begin{figure}[t]
\begin{lstlisting}
H_CLASS_ERGODIC = """- **H** requires an AERIAL robot -- the H behavior is wide-area coverage and that needs altitude. H robots run a persistent ergodic coverage policy: a stochastic transition matrix samples the next region, so over time they spend time in each region proportional to its importance. This policy is AUTONOMOUS and it helps explore an unknown environment. You do not tell an H robot where to go and you do not give it tasks. You choose only which robots are H. """
\end{lstlisting}
\caption{An example multi-agent protocol description for ergodic coverage.}
\label{fig:ma_protocol}
\end{figure}

\subsection{Subtask Generation}
\label{subsec:dispatcher}

A foundation model, the \textit{orchestrator}, decides at each invocation how many and which robots to delegate to a multi-agent protocol and which subtasks the remaining robots should complete. Note that allocations are \textit{not} done by the orchestrator directly and will instead be allocated algorithmically. The prompt provided to the orchestrator has a structure as detailed below; the prompt contains a pre-determined, fixed portion and a variable portion that changes with each invocation. As part of the prompt, the orchestrator is asked to a) return the set of subtasks as well as their dependencies in a directed acyclic graph (DAG) structure and b) employ Chain-of-Thought (COT) reasoning; these requirements, as well as the final validation and feedback step, are motivated by the approach of \cite{ravichandran2025heterogeneous}.

\textbf{Prompt structure.} Each prompt combines fixed instructions and the key information of the current state. The fixed instructions contain: a mission description in natural language; objectives, including competing priorities and rules; a team specification of each robot's platform and capabilities; task primitives with their required arguments; multi-agent protocols and their descriptions; rules for grounding subtasks in the map and reasoning about a changing environment; and an output schema requiring the orchestrator to return its reasoning, the robots assigned to protocols, the subtasks grounded in map regions, and their dependency graph.

The state is rebuilt at every invocation and contains: the semantic map, a short description of each region's relevant contents built from the team's observations, its time of last observation, and a summary of observation age across the map; robot state (location, role, current task); task failures since the previous invocation; and critical information like pending tasks, the number of unobserved regions, and staleness statistics (latest visit times, average data age, and the oldest-visited regions). Each prompt is therefore self-contained, and it does not grow by the accumulation of invocation history.

\textbf{DAG structure.} The orchestrator returns a list of robots delegated to protocols and a set of subtasks $\mathcal{T}_k$. Each subtask is a task primitive grounded in regions of the semantic map, and edges of a returned directed acyclic graph encode task precedence. 

\textbf{Chain-of-thought reasoning.} The output schema requires the orchestrator to write its reasoning before its decisions: which regions are relevant, pending tasks, and the reason behind the chosen role split and subtasks follow. 

\textbf{Validation and feedback.} Each subtask is parsed and checked against the task primitives and the current map: the primitive must exist, its arguments must match, and every region it names must exist in the semantic map. The dependency graph is checked for references to rejected, undefined subtasks, and cycles. Rejected outputs are returned to the orchestrator with a reasoning, allowing the planner to try again, up to a fixed number of attempts.

\subsection{Subtask Assignment} \label{sec:assignment}
% \victoria{Add the actual opt problem here}
Subtasks are assigned by a linear program minimizing the total time for ground robots to reach and complete their assigned subtasks. Let $M_k$ be the ground robots, $T'_k$ the subtasks whose DAG predecessors are complete, and $C_{nt}$ the shortest-path time for robot $n$ to reach and complete subtask $t$ over the believed traversable subgraph $R_k$ ($C_{nt} = \infty$ if $n$ cannot execute $t$'s behavior). The LP solves $\min_x \sum_{n,t} C_{nt} x_{nt}$ subject to $x_{nt} \in \{0,1\}$ and at-most-one assignment per robot and per task. The assignment is re-solved at every invocation allowing a subtask in progress to be reassigned to a robot with a lower time cost, unless prevented by the task requirements. A switching penalty prevents a swap for marginal differences. Unassigned subtasks remain candidates at the next iteration. The orchestrator may cancel a subtask it issued on incomplete information, such an observation that is later determined false. 
% Ground robots execute their assigned behavior by planning a shortest path over the believed traversable subgraph using Dijkstra's algorithm, with edge costs reflecting traversal time.

\subsection{Task and Event-Triggered LLM Prompting} \label{sec:triggers}
The orchestrator is re-invoked on several conditions, so that a plan written under partial observability stays valid in a time-varying environment. The signals are described below.

\subsubsection{Task Failure}
A robot cannot complete an assignment. 

\subsubsection{Change in Semantic Map}
The semantic map changes for an object of interest, such as new evacuees appearing for search-and-rescue. The user declares which kinds of objects are relevant and which events should trigger a response.

\subsubsection{Task Completion}
All robots have finished their subtasks. This allows for allocation of a pending subtask or re-delegation to multi-agent protocols.

\subsection{Feedback Structure} \label{sec:feedback}

In past work, content from all previous interactions with the LLM is included in the prompt; even though this can be lengthy, context windows typically are large enough \cite{ravichandran2025heterogeneous}. However, LLMs tend to do worse if too much context is provided \cite{generative-agents}; higher token usage means higher costs, and including all interactions means the length grows unbounded in continuous operation. Therefore, we take the Markovian stance to summarize the system well enough to not depend on the entire history. We include information about the current tasks, current robot status, map information for regions with relevant objects, and statistics on the age of observations.

\section{EXPERIMENTAL SETUP}

% \victoria{We should argue in this section that MANY other works have already tested this in real scenarios; our focus is on the textual interface scaling, not sim-to-real.}

We evaluate the framework in simulated search-and-rescue scenarios for disaster response, modeling spreading hazards, groups of people who must be evacuated, congestion, and blocked areas. Fire ignites regions randomly and spreads to neighboring regions over time. Blocked and burning regions cannot be entered by ground robots, and congestion temporarily slows them. However, aerial vehicles can view these areas from above, creating a heterogeneous mix of capabilities and information asymmetry. Groups of people appear, move along the street network, and must be escorted to a safe zone before a deadline or fire reaches them. We describe the setup of our experiments and comparisons below; results are discussed in the next section, Sec. \ref{sec:results}. To test the high-level planning capabilities in complex missions with hazard spread, large search region, and robot team size, we test this in simulation only. Prior work suggests this method can bridge the sim-to-real gap ~\cite{ravichandran2025heterogeneous}; we defer such a study to future work. 

\begin{figure}
    \vspace{0.25cm}
    \centering
    \includegraphics[width=1\linewidth]{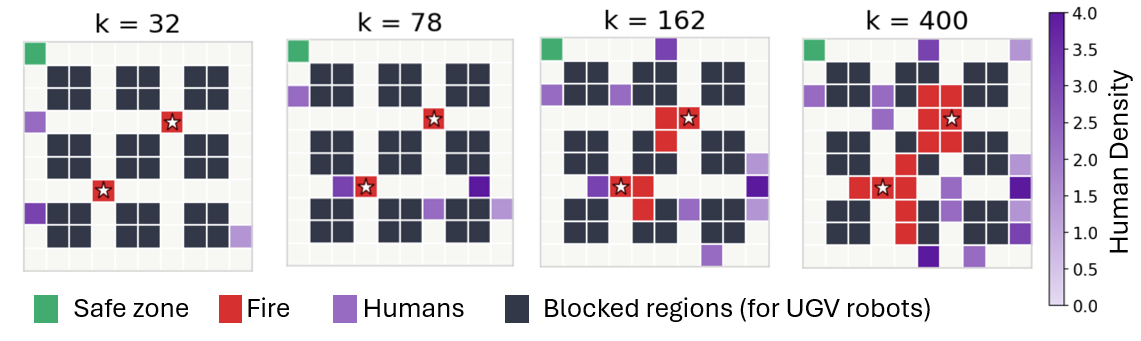}
    \caption{A $10\times10$ urban environment for disaster response, shown at selected steps. Dark regions are city blocks, which ground robots cannot enter; the remaining regions form the street network. Fire ignites at two regions (red blocks with stars) and spreads to neighbors over the course of the mission. Groups of people appear at street regions, vary in density, and move between steps; they must be escorted to the safe zone in the upper left (green). Robots are omitted for clarity. }
    \label{fig:Robot-env-no-robotr}
\end{figure}

\begin{table}[t]
\centering
\caption{Robot properties.}
\label{tab:sim_params}
\begin{threeparttable}
\begin{tabular}{llcc}
\hline
\makecell{Vehicle\\type} & Movement & \makecell{Speed\\(m/s)} &
\makecell{Sensor\\Range (m)} \\
\hline
Aerial (UAV) & survey      & 4.2 & 150  \\
             & go\_to, local\_search      & 2.9 & 50  \\
             & escort      & 1.4 & 50  \\
\hline
Ground (UGV) & go\_to, local\_search         & 2.9 & 50  \\
             & escort         & 1.4 & 50  \\
\hline
\end{tabular}
\begin{tablenotes}[flushleft]\footnotesize
\item A region is $250\,\mathrm{m}\times250\,\mathrm{m}$ and one timestep is $k=100\,\mathrm{s}$. Speeds enter the simulation as steps per region, the time a robot needs to cover one region: a drone sweeps it in a single pass, while a ground robot must drive the streets inside an urban block. Escort speed is set by the person on foot, likewise along streets.
\end{tablenotes}
\end{threeparttable}
\end{table}

\textbf{Setting.} The main scenario is a $10\times10$ street grid of
regions $\mathcal{R}$, $|\mathcal{R}|=100$, shown in Fig.~\ref{fig:Robot-env-no-robotr}. The simulation runs for a total of $K=500$ steps. Each region covers $250\,\mathrm{m}\times250\,\mathrm{m}$, and one simulation step is $\Delta k=100\,\mathrm{s}$ of mission time. Black regions are city blocks and are impassable to ground robots; the rest form the street network. Groups of humans appear over the first 400 steps as a Poisson process with rate $\lambda = N/K = 0.035$ groups per
step, for $N=14$ groups total. Group of humans $i$ arrives at a region drawn uniformly at random from $\mathcal{R}$, with occupancy $m_i$ (number of humans in the group) drawn uniformly from $\{1,\dots,4\}$. Each sampled group is assigned a route through the street network, so a group may relocate while the robot team is en route to its last reported position. Fire ignition and spread are held fixed across scenarios to maintain a controlled study; we instead vary the initial robot locations and the safe zone across two test cases.
% \victoria{The movements listed in the table are not explained until the task primitives paragraph later. I think this order is best for presentation, but there should be a brief explanation of what the movements mean in terms of robot speeds and that the purpose of each primitive and movement wil be explained later.}.
% \victoria{Since this has been renamed Robot Capabilities from Robot Parameters, let's also introduce the types of robots we're moving (this is in the table but could have a brief mention here too).}

\textbf{Task primitives.} Ground robots have four available task primitives: \texttt{go\_to} travels to a region, \texttt{inspect} observes a region at close range, \texttt{escort} carries people from a region to a safe zone, and \texttt{local\_search} searches the regions within a given radius of a last-known location. Each primitive's description states its arguments (e.g., \texttt{escort} needs the target and its location). 

\textbf{Robot Capabilities.}  Movement capabilities and sensing range are provided in Table \ref{tab:sim_params} and are based on commercially available platforms for drones and legged robots \cite{dji_consumer_drones, aerovision_survey_drones, unitree_go2}. A 250m x 250m area can cover multiple neighborhood blocks, so the sweep is equivalent to searching multiple streets. Congestion slows down the time to traverse a region for ground robots, so the robots compute via LP the time cost to reach destinations based on the semantic graph information. The time to traverse the regions during an escort task is bounded by the average human's speed of 1.4 m/s.

 At step $k$, robot $n$ at region $p_{n,k}$ observes the regions within its sensing range (see Table \ref{tab:sim_params}). Observations are written into the shared semantic map and are asymmetric across the team. Aerial robots cover a wide footprint but observe from altitude, so occupancy counts are reported with error or may be missed entirely; a miss is reported as "clear", which lowers the region's value to the coverage policy and suppresses revisits. Ground robots can observe at close range, giving them greater detail, like target identity, compared to those robots whose observations contain more noise.

\textbf{Multi-agent protocols.}
We provide the orchestrator access to a multi-agent coverage and exploration protocol that probabilistically determines which regions agents should visit next based on past collected data \cite{mendoza2026-dec-resilient-ergodic}. The orchestrator only needs to select which agents should adhere to this protocol, which can be variable across iterations. The multi-agent protocol then operates autonomously and, in our case, shares relevant information and updates the policy of the coverage and exploration task without invoking the orchestrator.

\textbf{Metrics}
We collect data across three scenarios with human appearances sampled as above; there is a time limit to the scenario to capture time-criticality. Results are summarized in Table \ref{tab:results} and will be discussed in the next section. The reported metrics are:
\begin{itemize}
    \item \textbf{Time to rescue (k)}: The number of time steps between a human's appearance and the time it arrives at the safe zone, averaged over all rescued humans.
    \item \textbf{Rescues (\%)}: The percentage of humans that are escorted to the safe zone.
    \item \textbf{\# Orchestrator calls}: Number of calls to the LLM.
    \item \textbf{Total tokens $(\times 10^6)$}: The total token count used for the mission; this includes input and output tokens where output tokens are split between internal "thinking" effort and the textual output.
\end{itemize}

% \textbf{Implementation Details}

\begin{table*}[t]
\centering
\begin{threeparttable}
\small
\setlength{\tabcolsep}{3pt}
\vspace{0.2cm}
\caption{Experimental results across configurations. Values are mean~$\pm$~std.}
\label{tab:results}
\begin{tabular}{cl lll cccc}
\toprule
& & \multicolumn{3}{c}{\textbf{Configuration}} & \multicolumn{4}{c}{\textbf{Results}} \\
\cmidrule(lr){3-5}\cmidrule(lr){6-9}
& Name & \makecell[l]{History\\Repr.} & \makecell[l]{Multi-Agent\\Protocols} & \makecell[l]{Re-prompt\\Trigger} &
\makecell{Time to\\rescue (k) $\downarrow$} & \makecell{Rescues\\(\%) $\uparrow$} & \makecell{ \# Orchestrator\\calls $\downarrow$} & \makecell{Total\\tokens ($\times10^6)$ $\downarrow$} \\
\midrule
% --- Comparison / baselines ---
\multirow{3}{*}{\rotatebox[origin=c]{0}{Comparisons}}
 & SPINE-HT \cite{ravichandran2025heterogeneous}    & Full  & No  & Task  & $106.9\pm23.8$ & $89.4\pm12.8$ & $33.6\pm12.5$ & $2.40\pm1.25$ \\
 & CoMuRoS \cite{borate2026comuros}  & Full & No  & Event    & $90.1\pm35.8$ & $32.5\pm19.0$ & $14.2\pm6.1$ & $0.45\pm0.35$ \\
 & Ours      & Summ. & Yes & All & $102.3\pm7.7$ & $98.1\pm2.8$ & $32.4\pm6.1$ & $0.41\pm0.07$\\
\midrule
% --- Ablations (fixed model) ---
\multirow{3}{*}{\rotatebox[origin=c]{0}{Ablations}}
 & History & Full & Yes & All & $105.2\pm15.0$ & $83.1\pm16.3$ & $25.8\pm3.6$ & $1.16\pm0.27$ \\
 & Re-Prompt  & Summ.    & Yes & Cadence   & $137.3\pm17.4$ & $95.6\pm4.7$ & $10.0\pm0.0$ & $0.14\pm0.01$ \\
 & Protocol Use   & Summ.    & No & All & $82.1\pm10.5$ & $93.1\pm10.5$ & $62.0\pm20.7$ & $0.89\pm0.28$ \\
\midrule
% --- Model variation (best config) ---
\multirow{2}{*}{\rotatebox[origin=c]{0}{\makecell{Model\\Variations}}}
 % & Ollama & Summ. & Yes & All & Ollama? & $\pm$ & $\pm$ & $\pm$ & $\pm$ \\
 & GPT-5.6 Luna & Summ. & Yes & All  & $109.7\pm18.9$ & $91.9\pm8.1$ & $26.6\pm2.1$ & $0.18\pm0.02$ \\
  & GPT-5.6-Terra & Summ. & Yes & All  & $106.1\pm15.2$ & $96.9\pm3.8$ & $27.2\pm2.0$ & $0.17\pm0.01$\\
  % & GPT-5.6-Terra (R=High) & Summ. & Yes & All  & $107.1\pm62.6$ & $96.2\pm5.1$ & $27.4\pm2.5$ & $0.19\pm0.02$ \\
 % \\
\bottomrule
\end{tabular}
\begin{tablenotes}[flushleft]\footnotesize
\item All runs in comparison and ablations use the Claude Sonnet 5.0 model with High Reasoning.
\item For each metric, we provide the mean and standard deviation.
\end{tablenotes}
\end{threeparttable}
\end{table*}

\section{Results and Discussion} \label{sec:results}

All experiments were run on a ThinkPad Lenovo laptop (Intel Core i7, 32\,GB RAM) and the simulator was written in Python\footnote{The GitHub repository will be available in the final submission}. Results are summarized in Table \ref{tab:results} across five runs (three with the first test case and two with the second).

\subsection{Comparison to Baselines} \label{sec:baseline}
We compare against two approaches with similar mission profile and architecture, SPINE-HT \cite{ravichandran2025heterogeneous}, which our framework builds on, and CoMuRoS \cite{borate2026comuros}. While both provide examples of search and rescue operations, neither operates continuously in a time-varying setting. Our larger but physics-extracted setting does not connect to ROS as these methods are implemented for, so we do not directly use their code and instead port the mechanisms that distinguish them from our approach: history representation, invocation (trigger) mechanism, and the absence of multi-agent protocols. An additional exploration task primitive is provided that allows the LLM to specify a sequence of regions to explore. CoMuRos also has capabilities such as LLM-determined relevance of observations and a direct chat interface, which we remove for direct comparison in this scenario.

CoMuRoS takes a hierarchical approach in which a centralized Task Manager LLM allocates subtasks, and each robot runs a local LLM for execution and classification of detected changes. In \cite{borate2026comuros}, an event callback is triggered by an observation that is determined by this robot's local LLM to be relevant to the mission.  Relevant events, task failures, and human instructions trigger replanning with the chat history supplied as context. 

This combination is ineffective in our setting where persistent search is necessary. This approach yields the fewest invocations of any configuration, $14.2$ against our $32.4$, while not reducing token usage due to the lack of multi-agent protocol delegation and a much worse rescue rate at $32.5\%$. Qualitatively, we see in the runs that CoMuRoS often chooses to allocate the faster aerial vehicles to any needed escort, temporarily neglecting exploration in favor of lower rescue time. While event-based triggering is faster at handling change, unexpected scenarios, and new data, it lacks the opportunity that switching on task completion provides for role adjustment. This greedy action devolves into a system that never updates because it is not proactively searching for relevant changes. In the Chain-of-thought reasoning, we consistently see it state that no exploration is needed because all regions have been observed once. Even with the alternative prompt ``There is an active fire spreading in the area. Humans are evacuating their houses and need to be escorted to the triage point.", performance was still poor.

\subsection{Ablation on Use of Multi-Agent Protocols} \label{ablation-protocols}
The purpose of providing access to multi-agent protocols is to allow the LLM to \textit{delegate} some agents to a well-defined group task that can operate autonomously, potentially continuously, freeing up LLM reasoning to focus on fewer tasks and agents. LLMs have been shown to degrade in capabilities when overloaded with too much information, requiring mechanisms to filter relevant information \cite{chen2024scalable}, so this approach allows for more resources to be allocated to the semantically challenging pieces of the mission that the LLM is more specialized to handle. We compare two approaches in an ablation study removing our multi-agent protocol that delegates coverage to an ergodic planner (see Table~\ref{tab:results}).
Ablating the multi-agent protocols means that the orchestrator must also handle the coverage by allocating exploration tasks directly. The orchestrator is now asked to provide a coverage route with a set of regions to visit. The coverage route is validated per robot against the class assignment and the map, and a rejected route is returned to the model as feedback for regeneration up to three times. For example, a robot whose route is missing or invalid keeps the one it was last given to avoid crashes due to typos or parsing errors from the output. Results show a decrease in the percent of rescues (~93.1\% vs 98.1\%) and almost twice the number of calls and token usage.

\subsection{Impact of Structure and Frequency of Prompting}
\begin{figure}
    \vspace{0.1cm}
    \centering
    \includegraphics[width=1.\linewidth]{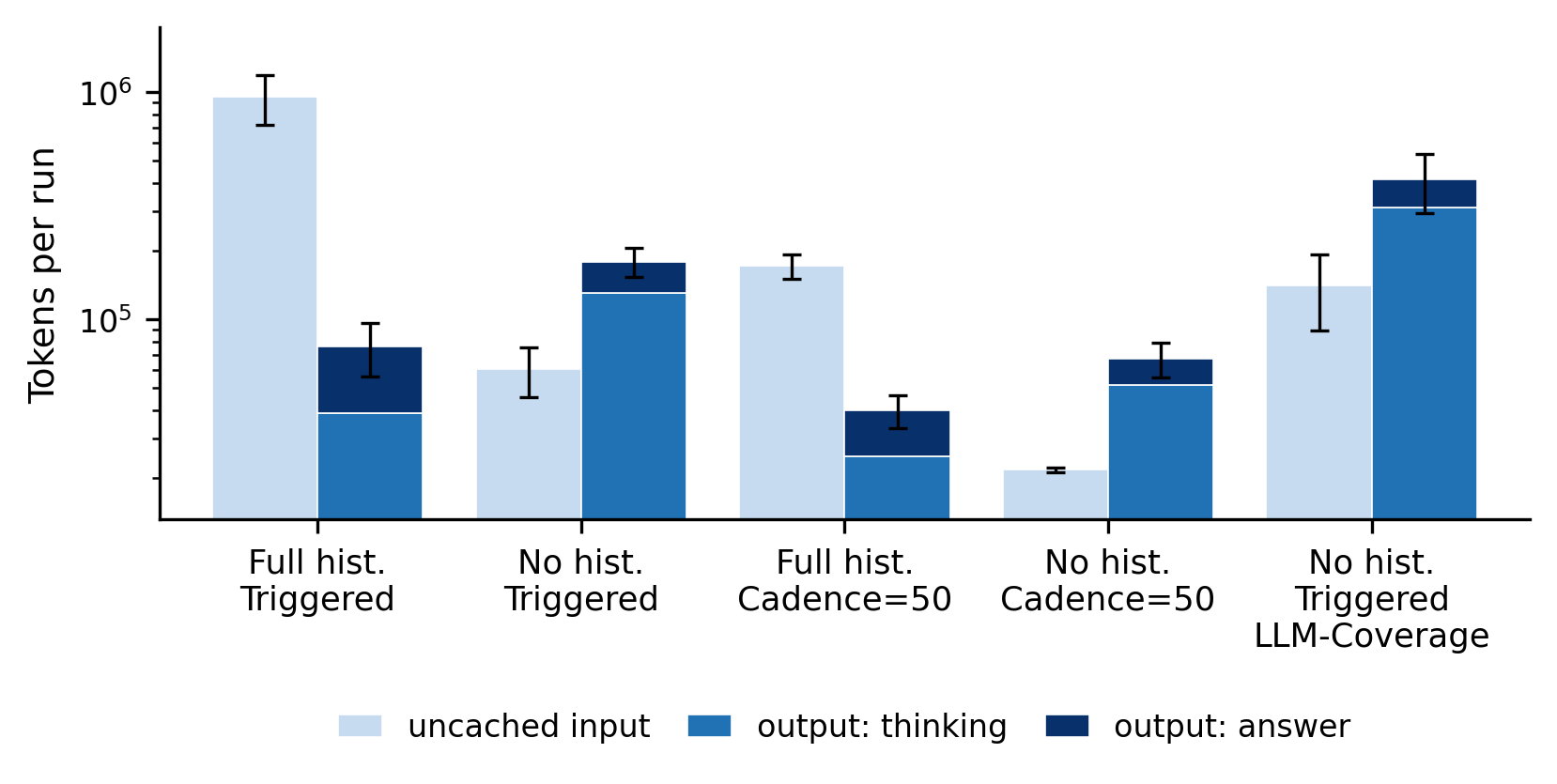}
    \caption{Total tokens used across ablations. We show that for the same triggering mechanism, the stateless prompt (which does not re-send the prompt history) uses fewer overall tokens though more output tokens. These output tokens come from additional "thinking" tokens the LLM needs.}
    \label{fig:token_usage_bar}
\end{figure}

\begin{figure}
    \centering
    \includegraphics[width=1.0\linewidth]{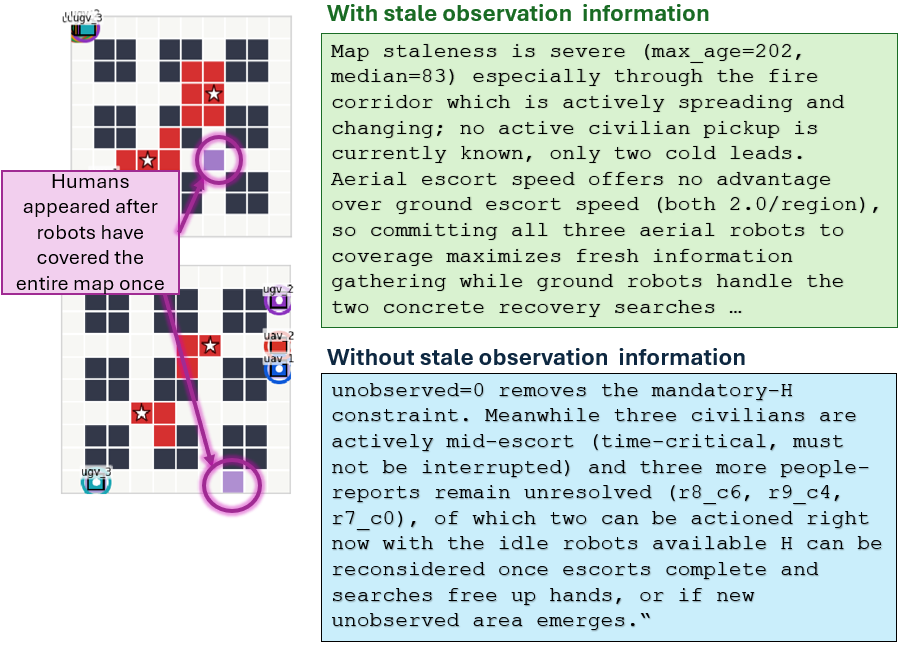}
    \caption{Effect of exposing observation staleness to the orchestrator.  With staleness in the prompt (green), the orchestrator sees $\max\text{age}=202$ and a median of 83, assigns all three aerial robots on coverage, and re-observes the region later in the run, recovering the last group of humans. Without it (blue), coverage looks complete ($\text{unobserved}=0$), the mandatory-H constraint is dropped, aerial robots are released to other work, and the group is never found and the map is only covered once.}
    \label{fig:data_age}
\end{figure}

In this set of analyses, we study how to adjust the prompt and prompt timing to improve outcomes. In previous work, the full prompt history or a sliding window \cite{chen2024scalable} is provided for context at each iteration. The length of the full history increases with call number, which loosely correlates with time step, and a sliding window risks leaving out relevant information. We show reduced token usage by instead providing a summary of history without degrading performance. It has been successful because we can represent the system with a Markovian state representation. For the same triggering scheme, "Summ." uses much fewer tokens than "Full" with similar or better levels of success (see Fig. \ref{fig:token_usage_bar}). Additionally, the LLM struggles to infer the importance of revisiting regions or collecting new data in the time-varying setting. We find that after providing observation statistics including average age and oldest region observations, the LLM does decide to continue exploring. Fig. \ref{fig:data_age} shows a case study of this impact. 

We also compare re-prompting triggers. In Sec. \ref{sec:baseline}, we compare the event- and task-based triggering mechanisms on the baselines. An MPC-style approach, where replanning is triggered at a set frequency, can lead to lower token usage with the same performance compared to our method. However, in addition to a slightly lower success rate, we see that the response time to events of interest is lower, which is a drawback in time-critical settings.

\subsection{Effect of Full History vs State Summary}

We observed that appending the full history costs $1.16e6$ tokens per run with a rescue success of $83.1\%$; summarizing using our approach reduces the cost to $0.41e6$ with a much higher success at $98.1\%$ (Table~\ref{tab:results}). We have previously discussed how a longer context may confuse instead of inform, which is supported by this data.

Fig.~\ref{fig:token_usage_bar} shows a further breakdown of token distribution. With full history the run is dominated by uncached input whereas without history, specifically the "thinking" portion of output tokens dominates, and thinking is proportionally more of the no history output usage. We interpret this switch as the orchestrator using more of its budget on reasoning about the mission instead of just re-reading the prompt.

\subsection{Model Variations}
All base experiments use Sonnet 5.0. To test whether the architecture can work across multiple models, we run our approach using GPT-5.6 Luna and GPT-5.6 Terra, holding scenario, seed, mission, and prompt fixed. We see a small degradation with Luna with percent success still above 90\%. All three stay within one standard deviation of each other on time to rescue (Table~\ref{tab:results}), suggesting compatibility across models with reasonable reasoning capabilities.

\section{CONCLUSION}

In this work, we introduce a framework for multi-agent orchestration in time-varying and continuous settings that leverages the advanced capabilities of grounded generative intelligence. Our approach improves scalability by managing prompt length and allowing the LLM to delegate to autonomous multi-agent protocols, freeing up reasoning. We incentivize persistent operation by improved re-prompting and feedback strategies. Together, these methods reduce token usage while improving the number of successful rescues and rescue speed in a search-and-rescue scenario. Opportunities for future work include orchestration using a distilled model as large, online models may not be available when infrastructure access is limited due to damage from a disaster. Locally centralized operation as a mix between our centralized approach and decentralized multi-agent algorithms may be necessary to scale to large distance with short-range robot communication. Finally, supporting human-provided input and mid-mission changes would broaden practical use.

\addtolength{\textheight}{-15cm}   % This command serves to balance the column lengths
                                  % on the last page of the document manually. It shortens
                                  % the textheight of the last page by a suitable amount.
                                  % This command does not take effect until the next page
                                  % so it should come on the page before the last. Make
                                  % sure that you do not shorten the textheight too much.

%%%%%%%%%%%%%%%%%%%%%%%%%%%%%%%%%%%%%%%%%%%%%%%%%%%%%%%%%%%%%%%%%%%%%%%%%%%%%%%%

%%%%%%%%%%%%%%%%%%%%%%%%%%%%%%%%%%%%%%%%%%%%%%%%%%%%%%%%%%%%%%%%%%%%%%%%%%%%%%%%

\section*{ACKNOWLEDGMENT}

The authors acknowledge the use of Claude Opus for partial code generation in experimental evaluation.

%%%%%%%%%%%%%%%%%%%%%%%%%%%%%%%%%%%%%%%%%%%%%%%%%%%%%%%%%%%%%%%%%%%%%%%%%%%%%%%%

\bibliographystyle{IEEEtran} 
\bibliography{ref} 

% \newpage
%%%%%%%%%%%%%%%%%%%%%%%%%%%%%%%%%%%%%%%%%%%%%%%%%%%%%%%%%%%%%%%%%%%%%%%%%%%%%%%%

\end{document}